\global\def\draftcontrol{0}

   \def\versionno{top string review}

\catcode`\@=11

\expandafter\ifx\csname draftcontrol\endcsname\relax\global\def\draftcontrol{0} 
\fi 

{\count255=\time\divide\count255 by 60 
\xdef\hourmin{\number\count255} 
\multiply\count255 by-60\advance\count255 by\time 
\xdef\hourmin{\hourmin:\ifnum\count255<10 0\fi\the\count255}} 
\def\draftdate{\number\month/\number\day/\number\year\ \ \ \hourmin } 

\newcommand\makepapertitle{\par

  \begingroup 
    \renewcommand\thefootnote{\@fnsymbol\c@footnote}%
    \def\@makefnmark{\rlap{\@textsuperscript{\normalfont\@thefnmark}}}%
    \long\def\@makefntext##1{\parindent 1em\noindent 
            \hb@xt@1.8em{%
                \hss\@textsuperscript{\normalfont\@thefnmark}}##1}%
     \newpage 
     \global\@topnum\z@   
     \@makepapertitle 
     \thispagestyle{empty}\@thanks 
  \endgroup 
  \setcounter{footnote}{0}%
  \global\let\thanks\relax 
  \global\let\makepapertitle\relax 
  \global\let\@makepapertitle\relax 
  \global\let\@thanks\@empty 
  \global\let\@author\@empty 
  \global\let\@date\@empty 
  \global\let\@title\@empty 
  \global\let\title\relax 
  \global\let\author\relax 
  \global\let\date\relax 
  \global\let\and\relax 
  \def\version{\let\version\@version\@gobble} 
} 
\def\@makepapertitle{%
  \newpage 
   \ifnum\draftcontrol=1 {} 
   \version\versionno 
   \vskip 5.5em%
   \else 
   \hfill\hbox to 3cm {\parbox{4cm}{\@pubnum}\hss}%
   \vskip 6.5em%
   \fi 
   \begin{center}%
   \let \footnote \thanks 
      {\hskip -0\textwidth \hbox to 1\textwidth%
        {\centerline{\Large\bf{\noindent\@title}}}}%
     \vskip 2em%
     {\normalsize
       \lineskip .5em%
       \begin{tabular}[t]{c}%
         \@author 
       \end{tabular}\par}%
     \vskip 1.5em%
     {\@bstract}%
     \end{center}%
     \vfill
     \@date%
     \vskip 1.5em%
   \par 
} 

\gdef\@pubnum{} 
\def\pubnum#1{%
  \gdef\@pubnum{#1}} 

\gdef\@bstract{} 
\def\Abstract#1{%
  \gdef\@bstract{%
   \parbox{\textwidth-0pc}{%
   \centerline{\bf Abstract}\penalty1000 
   \noindent
   \renewcommand\baselinestretch{1.0} 
   {#1}}} 
} 

\gdef\@email{}
\def\email#1{%
   \gdef\@email{%
   Email: {\tt #1}}
}

\def\ps@paper{\let\@mkboth\@gobbletwo%
     \ifnum\draftcontrol=1 
        \def\@oddfoot{\hbox to \textwidth{\tiny \versionno \hfil\tiny\draftdate}%
        \hskip -\textwidth \hbox to \textwidth{\hfil\rm\thepage\hfil}}%
     \else\def\@oddfoot{\hbox to \textwidth{\hfil\rm\thepage\hfil}} 
     \fi 
     \let\@evenfoot\@oddfoot 
} 

\def\body{\clearpage 
          \pagestyle{paper} 
        } 
\newenvironment{acknowledgments}{%
\vskip 3.25ex 
\addcontentsline{toc}{section}{Acknowledgments}
\noindent {\bf Acknowledgments} 
} 

\def\@version#1{\ifnum\draftcontrol=1 
\typeout{}\typeout{#1}\typeout{} 
\vskip3mm\centerline{\hbox{\fbox{\normalsize{\tt DRAFT -- #1 -- } 
                   {\draftdate}}}}\vskip3mm 
\fi} 
\let\version\@version 
\long\def\eqlabel#1{\ifnum\draftcontrol=1 
                    \tag@false  
                    \tag*{(\theequation) \hbox to -0.2cm{\hspace{0cm}\small{#1}\hss}} 
                    \refstepcounter{equation}  
                    \edef\@currentlabel{\theequation} 
                    \ltx@label{#1}          
                    \else 
                    \label{#1} 
                    \fi 
                    } 
\let\st@bibitem\@bibitem 
\let\st@lbibitem\@lbibitem 
\ifnum\draftcontrol=1 
  \def\@bibitem#1{%
    \st@bibitem{#1}\a@@label{#1}\ignorespaces} 
  \def\@lbibitem[#1]#2{%
    \st@lbibitem[#1]{#2}\a@@label{#2}\ignorespaces} 
  \def\a@@label#1{%
    \gdef\a@lab{\smash{\normalfont\small#1}} 
    \ifvmode 
      \if@inlabel 
        \global\setbox\@labels\hbox{%
          \llap{\a@lab\let\a@lab\relax 
                \kern\@totalleftmargin\kern\marginparsep}%
          \box\@labels}%
      \fi 
    \fi} 
\fi 

\documentclass[12pt,letterpaper]{article} 

\usepackage{amsmath,bm,amsfonts,amssymb,array,calc,amsthm,rotating}
\usepackage{epsfig,psfrag} 
\usepackage{graphicx}
\usepackage{color}
\usepackage{float}
\usepackage[colorlinks=false]{hyperref}
\renewcommand\baselinestretch{1.25} 
\renewcommand\section{\@startsection {section}{1}{\z@}%
                                   {-3.5ex \@plus -1ex \@minus -.2ex}%
                                   {2.3ex \@plus.2ex}%
                                   {\normalfont\large\bfseries}} 
\renewcommand\subsection{\@startsection{subsection}{2}{\z@}%
                                   {-3.25ex\@plus -1ex \@minus -.2ex}%
                                   {1.5ex \@plus .2ex}%
                                   {\normalfont\normalsize\bfseries}} 
\renewcommand\subsubsection{\@startsection{subsubsection}{3}{\z@}%
                                   {-3.25ex\@plus -1ex \@minus -.2ex}%
                                   {1.5ex \@plus .2ex}%
                                   {\normalfont\normalsize\it}} 
\renewcommand\paragraph{\@startsection{paragraph}{4}{\z@}%
                                   {-3.25ex\@plus -1ex \@minus -.2ex}%
                                   {1.5ex \@plus .2ex}%
                                   {\normalfont\normalsize\bf}} 
\renewcommand\subparagraph{\@startsection{subparagraph}{5}{\z@}%
                                   {-1.25ex\@plus -1ex \@minus -.2ex}%
                                   {0ex \@plus .2ex}%
                                   {\normalfont\normalsize\it}}

\numberwithin{equation}{section}

\long\def\@makecaption#1#2{%
  \vskip\abovecaptionskip
  \sbox\@tempboxa{{\bf #1:} #2}%
  \ifdim \wd\@tempboxa >\hsize
    {\small\bf #1:} {\small #2}\par
  \else
    \global \@minipagefalse
    \hb@xt@\hsize{\hfil\box\@tempboxa\hfil}%
  \fi
  \vskip\belowcaptionskip}

\renewcommand*\l@section[2]{%
  \ifnum \c@tocdepth >\z@
    \addpenalty\@secpenalty
    \addvspace{.5em \@plus\p@}%
    \setlength\@tempdima{1.5em}%
    \begingroup
      \parindent \z@ \rightskip \@pnumwidth
      \parfillskip -\@pnumwidth
      \leavevmode \bfseries
      \advance\leftskip\@tempdima
      \hskip -\leftskip
      #1\nobreak\hfil \nobreak\hb@xt@\@pnumwidth{\hss #2}\par
    \endgroup
  \fi}
\renewcommand*\l@subsection{\addvspace{.0em \@plus\p@}\@dottedtocline{2}{1.5em}{2.3em}}
\renewcommand*\l@subsubsection{\addvspace{-.2em \@plus\p@}\@dottedtocline{3}{3.8em}{3.2em}}

\definecolor{refcol}{rgb}{0.2,0.2,0.8}
\definecolor{eqcol}{rgb}{.6,0,0}
\definecolor{purple}{cmyk}{0,1,0,0}

\gdef\@citecolor{refcol}
\gdef\@linkcolor{eqcol}
\def\colorlinkspurple{\gdef\@urlcolor{purple}}
\def\colorlinksblue{\gdef\@urlcolor{blue}}
\def\colorlinksred{\gdef\@urlcolor{red}}

\def\ie{{\it i.e.}}

\def\cf{{\it cf.}}

\def\revise#1       {\raisebox{-0em}{\rule{3pt}{1em}}%
                     \marginpar{\raisebox{.5em}{\vrule width3pt\ 
                     \vrule width0pt height 0pt depth0.5em 
                     \hbox to 0cm{\hspace{0cm}{%
                     \parbox[t]{4em}{\raggedright\footnotesize{#1}}}\hss}}}}

\def\calc         {{\cal C}}

\def\calf         {{\cal F}}

\def\call         {{\cal L}} 
\def\calm         {{\cal M}} 
\def\caln         {{\cal N}}

\def\del          {\partial} 
\def\delbar       {\bar\partial}

\newcommand\topa[2]{\genfrac{}{}{0pt}{2}{\scriptstyle #1}{\scriptstyle #2}}

\def\sqr#1#2{{\vcenter{\vbox{\hrule height.#2pt   
 \hbox{\vrule width.#2pt height#1pt \kern#1pt 
 \vrule width.#2pt}\hrule height.#2pt}}}}

\renewcommand{\P}{\mathbb P}

\newcommand{\Z}{\mathbb Z}

\newcommand{\Fcal}{\mathcal F}

\newcommand{\Ocal}{\mathcal O}

\newcommand{\Ncal}{\mathcal N}

\newcommand{\ep}{\epsilon}

\newcommand{\beq}{\begin{equation}}
\newcommand{\eq}{\end{equation}}
\newcommand{\req}[1]{(\ref{#1})}

\newcommand{\epo}{\epsilon_1}
\newcommand{\ept}{\epsilon_2}

\catcode`\@=12 

\begin{document} 


\title{B-Model Approach to Instanton Counting}

\pubnum{
}
\date{December 2014}

\author{
Daniel Krefl$^{a}$ and Johannes Walcher$^{b}$ \\[0.2cm]
\it $^{a}$ Center for Theoretical Physics \\ 
\it Seoul National University,
Seoul, South Korea \\[.2cm]
\it $^{b}$ Department of Physics, and Department of Mathematics and Statistics \\ 
\it McGill University,
\it Montreal, Quebec, Canada}

\Abstract{
The instanton partition function of $\caln=2$ gauge theory in the general $\Omega$-background
is, in a suitable analytic continuation, a solution of the holomorphic anomaly equation known
from B-model topological strings. The present review of this connection is a contribution
to a special volume on recent developments in $\caln=2$ supersymmetric gauge theory and the 
2d-4d relation, edited by J.\ Teschner.
}

\makepapertitle

\body

\version\versionno

\vskip 1em

\section{Introduction and Key Ideas}

The instanton partition function of $\caln=2$ supersymmetric quantum field theories
\begin{equation}
\eqlabel{part}
Z^{\rm inst}(a,\epsilon_1,\epsilon_2;\Lambda)
\end{equation}
is of algebra-geometro-physical interest for at least three different, though related, 
reasons. First of all, by its very definition, $Z^{\rm inst}$ encapsulates the cohomology 
of the moduli space of instantons, supersymmetric solutions of the underlying classical 
field theory, and the algebraic structures on that space (chapter \cite{V3}). Secondly, within
the 2d-4d correspondences of Alday-Gaiotto-Tachikawa, and Nekrasov-Shatasvhili (chapter \cite{V9}), 
the instanton partition function connects supersymmetric field theories with 
the world of completely integrable systems and their quantization, specifically Hitchin
systems (see chapter \cite{V2}). Thirdly, $Z^{\rm inst}(a,\epsilon_1,\epsilon_2;\Lambda)$ 
contains information about the structure of the Coulomb branch that goes beyond the weakly 
coupled description in a Lagrangian field theory. After a suitable analytic continuation,
it allows to calculate interesting physical quantities everywhere in the moduli space of 
vacua and marginal couplings, and thereby to study a variety of dualities.

It is in fact, this latter aspect of the instanton partition function that is closest
to the approach pioneered by Seiberg and Witten for solving the low-energy
dynamics of $\caln=2$ supersymmetric field theories, by exploiting the global
constraints on the structure of the moduli space coming from special geometry and
modular invariance. The basic ideas are easily explained.

The instanton partition function $Z^{\rm inst}$ calculated via localization (see chapter
\cite{V3}) is a series
\begin{equation}
\eqlabel{scheme}
Z^{\rm inst} \sim \sum_n \Lambda^n R_n(a,\epsilon_1,\epsilon_2)\,,
\end{equation}
with rational functions $R_n$ of the $\Omega$-background parameters $\epsilon_{1,2}$,
and the Coulomb branch parameters $a$, that converges well for small instanton counting
parameter $\Lambda$. As explained by Nekrasov \cite{N02}, the Seiberg-Witten solution
for the low-energy effective action is recovered in the non-equivariant limit
$\epsilon_{1,2}\to 0$. Specifically, the $\caln=2$ prepotential is the residue
\begin{equation}
\eqlabel{limit}
\calf^{(0)}(a;\Lambda) = \lim_{\epo,\ept\to 0} \bigl(\epo\ept\log 
Z^{\rm inst}(a,\epo,\ept;\Lambda)\bigr)\,,
\end{equation}
after the perturbative expansion of the free energy for small $\Lambda$ (in the 
asymptotically free case, this is equivalent to the weak-coupling limit $a\to\infty$).
It coincides with the prepotential obtained from the Seiberg-Witten effective geometry
(the family of hyper-elliptic curves together with the differential), which
captures the low-energy dynamics and is the basis for the various embeddings into 
string theory. Most specifically, the $\calf^{(0)}$ appears in the so-called geometric 
engineering limit of the prepotential governing the compactification of type II string 
on a (non-compact) Calabi-Yau manifold \cite{kachruetal}.

It has been a natural question to ask for the analytic and geometric characterization of
the terms in $Z^{\rm inst}$ of higher order in $\epsilon_1,\epsilon_2$, and their
physical interpretation. As anticipated already by Nekrasov \cite{N02}, the answer
is most immediate on the special slice in coupling constant space $\epo=-\ept$.
By a detour in one higher dimension, one can see that in general, the $\Omega$-background 
in the gauge theory arises in the string/M-theory constructions from a vacuum 
expectation value of the gravi-photon field strength, of the form
\begin{equation}
F = \epsilon_1 dx_1\wedge dx_2 + \epsilon_2 dx_3\wedge dx_4
\end{equation}
(itself a limit of the Melvin background, or ``flux-trap'' \cite{simeon} in other duality
frames). The specialization $\epo=-\ept$ corresponds to a self-dual gravi-photon
background, and the expansion coefficients of the supersymmetric free energy in 
$\epo=-\ept=:g_s$ are identified with the higher-derivative F-term couplings 
$R^2 F^{2g-2}$ in the effective action \cite{AGNT93}, which can be computed as the
{\it topological} string genus-$g$ free energy \cite{BCOV93b}. In the geometric engineering
limit, one recovers the expansion of the instanton partition function.
\begin{equation}
\eqlabel{recover}
\log Z^{\rm inst}(a,g_s,-g_s;\Lambda) = \sum_{g=0}^\infty g_s^{2g-2}  \calf^{(g)}(a)
\end{equation}
A natural way to test this physical interpretation is to re-calculate the $\calf^{(g)}$
using the string theory methods. In the topological B-model, the most universal of these 
methods is the holomorphic anomaly of BCOV \cite{BCOV93b}. 

The basic message of the holomorphic anomaly method is that the higher order
corrections $\calf^{(g)}(a)$ can still be continued throughout moduli space,
in particular any strong coupling regions, but (in distinction to the prepotential 
$\calf^{(0)}$), they are no longer holomorphic functions of $a$. The physical origin 
of this non-holomorphicity are the infrared effects, degenerating Riemann surfaces
in the perturbative string theory \cite{BCOV93b}, or the distinction between
1 PI and Wilsonian effective action from the point of view of the field theory
\cite{shva}. Mathematically, the holomorphic anomaly is an expression of the
competition between holomorphy and modular invariance \cite{BCOV93b},
and can also be viewed as an embodiment of the wave-function nature of the
topological partition function \cite{W93}.

The holomorphic anomaly {\it equation} dictates the non-holomorphic dependence of 
$\calf^{(g)}(a,\bar a)$ recursively in the order of the expansion, $2g-2$. The
meromorphic function on moduli space that is thereby left undetermined at each order 
is known as the holomorphic ambiguity and can, under favorable circumstances, be
determined by imposing appropriate principal parts or ``boundary conditions'' at 
the various singular points.

It was shown by Klemm and Huang \cite{HK06} that the holomorphic anomaly commutes
with the geometric engineering limit, and can be used to completely recover the 
$\calf^{(g)}$ in the expansion \eqref{recover}. Even though a detailed derivation 
of the holomorphic anomaly from the gauge theory point of view is missing, the equation 
itself can be written down based solely on the special geometry data on the
moduli space that can be obtained from the prepotential $\calf^{(0)}$. And at 
least in all examples with low-dimensional moduli space, the boundary conditions 
at the monopole/dyon points are sufficient to completely fix the holomorphic
ambiguity, and thereby make the holomorphic anomaly ``integrable'' in that sense.

From the point of view of instanton counting \eqref{part}, this discussion of 
the holomorphic anomaly appears as rather tangential. After all, the holomorphy and
integrability of the higher order corrections is built into the formalism, while
the underlying spectral geometry is completely determined by the first order
classical term, \ie, the prepotential. There are nevertheless several very good
reasons to explore the connection further, and in particular, to understand the 
extension of the holomorphic anomaly to the full two-parameter $\Omega$-background, 
away from the specialization $\epo=-\ept$. 

From the gauge theory point of view, the precise role of the holomorphic anomaly, or
the wave-function nature of $Z^{\rm inst}$ is not completely understood, for instance 
in the context of the quantum integrable systems. Moreover, the possible calculation 
of $Z^{\rm inst}$ as a ``sum over instantons'' (or some other semi-classical 
configurations) around other points in moduli space remains to be explored. 
While conformal field theory in principle provides formal expressions for 
$Z^{\rm inst}$ in terms of certain contour integrals also elsewhere in moduli
space, these have been evaluated explicitly only in a limited number of situations.
The continuation of the $\calf^{(g)}$ to other points in moduli space via the 
holomorphic anomaly provides a very welcome benchmark for such calculations.

From the string theory point of view, the existence of the second deformation 
parameter itself is the most intriguing aspect. Indeed, while the role of $g_s
=\sqrt{-\epo\ept}$
as the genus-counting parameter, \ie, the topological string coupling
constant, is readily appreciated, the existence of a second ``string-coupling 
like'' parameter is much more mysterious. Since, in much more generality
than the restricted topological context, string theory does not have any
free parameters, the absence of a worldsheet description would be in tremendous
tension with the overall picture. To be sure, the role of the second parameter
from the macroscopic space-time, or M-theory point of view is completely clear,
see \cite{N02,HIV}, as well as the connection with refinement and
categorification \cite{GSV}. What is missing is the {\it microscopic} explanation.

The main point of the present contribution is to highlight the observation that with the right
choice of parameterization of the coupling constants, the deformation away from the 
special slice $\epo=-\ept$ is indeed as simple as it could be: The higher order 
corrections for general $\epo,\ept$ still satisfy the holomorphic anomaly equations,
with deformation only in the boundary conditions. In particular, a single 
infinitesimal coupling constant is sufficient. This was first pointed out in
\cite{KW10a,HK10,KW10b}. These methods therefore allow the calculation (via ``analytic''
continuation) of $Z^{\rm inst}$ around points in moduli space other than the weak
coupling regime. This constitutes a benchmark for testing the
2d-4d relation this special volume is about away from $\Lambda\rightarrow 0$.
Coming back to string theory, these observations have allowed the application of 
the holomorphic anomaly equation for the B-model calculation of refined BPS invariants 
of local Calabi-Yau manifolds \cite{KWun,HK11,WIP}. This can be viewed as further
evidence that the second parameter should be lifted to the topological string (not 
necessarily as a coupling constant, but rather as a deformation parameter),
and has been as well applied and interpreted in the context of quantum geometry and 
quantum integrability \cite{ACDKV11}.
Among the possible stringy explanations of the refinement, we will outline in 
somewhat more detail an intriguing relation to orbifolds and orientifolds, following  
\cite{KW10a,DK12b}.

Before closing this introduction, it seems worthwhile to emphasize once again that
in this Chapter, we are discussing the instanton partition function from the
point of view of the ``B-model'', meaning the global structure of the moduli space,
special geometry and modular invariance. In contrast to \eqref{scheme}, which is
exact in $\epo$, $\ept$, but perturbative in $\Lambda$, the B-model provides answers
that are exact in the instanton expansion, but perturbative in $\epo$, $\ept$.

\section{Geometric Engineering}

Large classes of supersymmetric gauge theories in various dimensions can be 
systematically obtained from string-, M- and F-theory compactifications. This is 
usually referred to as geometric engineering, as the geometry of the compactification 
manifold $X$ determines the effective gauge theory in the field theory limit. 
We only give a lightning overview, excellent pedagogical reviews being available in the 
canonical literature, see for instance, \cite{Klemm97,mayr}.

Any given gauge theory can typically be realized in several ways in string theory.
These different constructions are then related by various dualities and limiting
procedures. Hence, depending on the gauge theory to be investigated via geometric 
engineering, and the specific gauge theory property under investigation, a convenient 
duality frame has to be chosen. A common feature of all geometric engineering approaches 
is that in order to decouple string and gravity effects, the compactification manifold $X$ 
has to feature a local singularity, perhaps in the guise of a brane. 

We are interested in $\Ncal=2$ supersymmetric gauge theories in four dimensions, their
low-energy effective prepotential $\Fcal^{(0)}(a,m)$, and higher derivative F-term 
couplings. These are (modulo the holomorphic anomaly) holomorphic functions of the 
Coulomb moduli $a_i$ and masses of matter fields $m_i$, and receive their essential 
contributions from the space-time instantons. This class of theories can be conveniently 
engineered and investigated in a type IIA/B superstring framework by compactification
on a local (non-compact) Calabi-Yau 3-fold, which yields, under certain conditions, a 
four-dimensional $\Ncal=2$ supersymmetric gauge theory theory with decoupled gravity. 

To be specific, consider type IIA string theory compactified on a Calabi-Yau 3-fold
$X$. Four-dimensional abelian gauge fields arise in the Ramond-Ramond sector by 
dimensional reduction on the even cohomology of $X$. But since perturbative
string states do not carry Ramond-Ramond charge, in order to obtain interesting
non-abelian gauge groups, we must include non-perturbative effects. In particular,
D2-branes wrapped on the (compact) 2-cycles of the compactification geometry 
represent objects electrically charged under the corresponding abelian gauge
fields. The masses of these states being proportional to the K\"ahler 
class (volume) $t_{f_i}$ of the wrapped 2-cycles, one needs $t_{f_i}\rightarrow 0$ 
in order to have massless charged gauge bosons. 

In fact, it is best to view the Calabi-Yau compactification as the dimensional reduction 
of a K3 compactification near an ADE singularity. The gauge group originates in six 
dimensions from the (compact) homology of the singularity (of ADE type for ADE gauge 
group), while further dimensional reduction (on a copy, $B\cong \P^1$, of complex 
projective space to be specific) leads to an $\Ncal=2$ gauge theory in four dimensions. 
In this process, the bare gauge coupling $g_{{\it YM}}$ of the four-dimensional theory is 
proportional to the K\"ahler class $t_B$ of the 2-dimensional manifold used for the 
reduction, \ie, $t_B\sim 1/g^{2}_{{\it YM}}$. In order to decouple gravity (and 
stringy effects) it is sufficient to send the coupling constant to zero, since it pushes 
the string scale to infinity \cite{kachruetal}. This means that we are interested 
in the limit $t_B\rightarrow\infty$. 

In order to satisfy the Calabi-Yau condition, the compactification space can 
not be given by a direct product, but rather must have the structure of a fibration,  
\begin{equation}
F\rightarrow X\rightarrow B\,,
\end{equation}
where the fiber geometry $F$ (with the ADE singularity) determines the gauge group while 
the base geometry $B$ the effective gauge coupling in four dimensions. Note that the 
fibration structure also allows to incorporate matter content via local enhancement of 
the fiber singularity.

It is important to keep in mind that the limits $t_B\to\infty$ and $t_{f_i}\to 0$ of base 
and fiber K\"ahler classes are not independent. This can be illustrated best at hand of 
a concrete example. Consider the geometric engineering of pure $SU(2)$ along the lines 
sketched above, as originally discussed in \cite{kachruetal}. To obtain the two 
charged gauge bosons, W$^+$ and W$^-$, it is sufficient to fiber a $\P^1$ over the 
base $\P^1$. [The different ways this can be done are labeled by an integer and the 
corresponding geometries correspond to the Hirzebruch surfaces. The Calabi-Yau 3-fold
itself is the total space of the anti-canonical bundle over this complex surface.
All Hirzebruch surfaces give rise to pure $SU(2)$ in four dimensions.] Recall that 
in the weak coupling regime the running of the gauge coupling is given by
\begin{equation}
\frac{1}{g^2_{YM}}\sim \log\frac{m}{\Lambda}\,,
\end{equation}
where $m$ denotes the mass of the W-bosons and $\Lambda$ the dynamical scale. With
the above identifications, we learn that we have to take the limit in a way such that 
$t_B\sim \log t_f$ holds. The precise proportionality constant can be fixed as follows: 
We know that at weak coupling the instanton corrections to the bare gauge coupling 
go in powers of $\left(\Lambda/a\right)^4$, with $a$ the Coulomb modulus. 
Correspondingly, we have to scale $e^{-t_b}\sim \delta^4\, \Lambda^4$ and $t_f\sim 
\delta\, a$ as $\delta \rightarrow 0$, which constitutes the map between the string 
moduli and the gauge theory parameters, for pure $SU(2)$.

The useful property of the type IIA string construction is that the space-time instanton
corrections are mapped to world-sheet instanton corrections.  Qualitatively, this is 
clear from the relation between the 2-cycles and the gauge coupling and gauge bosons
sketched above: The Euclidean string worldsheet is wrapped around the 2-cycles of 
the geometry with worldsheet instanton action $S\sim d_b\, t_b+d_f\, t_f$, where 
$d_b$ and $d_f$ refer to wrapping numbers. In particular, this means that we do not 
have to consider the full type IIA string theory to investigate the gauge theory from 
a string point of view. Rather, the topological sector is sufficient, \ie, the 
topological string amplitudes which capture world-sheet instanton corrections. 

Starting from the topological string tree-level amplitude, taking the above
gauge theory limit yields the space-time instanton corrections to the gauge theory
prepotential. The higher-genus amplitudes encode the gravitational corrections,
as sketched in the introduction. In this way, the string theory provides both
a conceptual framework, and a host of computational methods to investigate
non-perturbative effects in supersymmetric gauge theories. 

There is, however, one important subtlety to keep in mind. Since the geometric 
engineering limit involves $t_{f_i}\rightarrow 0$, the compactification geometry is 
in fact singular, and we are not expanding the string amplitudes around the large 
volume point in moduli space. Hence, if we compute the topological string amplitudes 
using the usual A-model techniques, which are valid at large volume (such as, 
localization \cite{K94} or the topological vertex \cite{AKMV05}), these amplitudes 
have to be analytically continued before we can take the limit. 
[The necessity of this analytic continuation is quite clear already from the fact that 
the large volume expansion of the topological amplitudes is a series in the exponentiated
K\"ahler moduli, whereas the gauge theory prepotential at weak coupling is an expansion
into negative powers of the Coulomb branch parameter.]

For the topological string amplitudes of the $SU(2)$ engineering geometry sketched 
above, the analytic continuation can be achieved via a relatively simple resummation 
\cite{KMT02}. In more complicated examples, one has to switch to the B-model mirror
of the type IIA string background in order to perform the analytic continuation.
We recall that in general, mirror symmetry maps worldsheet instanton corrections to
the expansions of classical geometric quantities. For instance, tree-level worldsheet
instantons in type IIA are encoded in the period integrals of the type IIB mirror
geometry. As these periods can be calculated as solutions of simple linear differential
equations, their analytic continuation all over moduli space is straight-forward.

Under the geometric engineering limit along the lines reviewed above, mirror
symmetry may be seen as the stringy origin of the Seiberg-Witten solution of 
$\Ncal=2$ gauge theory. That is, the Seiberg-Witten curve and differential arise 
in the limit of the mirror Calabi-Yau threefold which is mirror dual to the type 
IIA engineering geometry. In particular, as its B-model parent, the Seiberg-Witten 
geometry naturally provides a global description of the moduli space. 

So far we mainly had in mind spherical world-sheet instantons yielding instanton 
corrections to the gauge theory prepotential. However, perhaps the most useful 
property of this stringy construction is that it allows to calculate gravitational 
corrections to the $\Ncal=2$ gauge theory, originating from world-sheet instantons 
of higher genus. In detail, the genus-$g$ topological string amplitude yields 
$R^2 F^{2g-2}$ corrections to the gauge theory \cite{AGNT93}, which can be calculated 
very efficiently via a specific topological string B-model technique, namely the 
holomorphic anomaly equation, all over the Coulomb moduli space. This is the
subject to which we now turn.

\section{B-model}

It has been observed some time ago in \cite{HK06} that the free energy of four 
dimensional $\Ncal=2$ supersymmetric gauge theory with gravitational corrections satisfies 
the holomorphic anomaly equations of \cite{BCOV93a,BCOV93b}. This can be seen as a 
consequence of the geometric engineering approach to $\Ncal=2$ gauge theories, where the 
gauge theory free energy follows as a specific limit of the topological string free energy 
on a corresponding engineering Calabi-Yau, as outlined in the previous section.

Since the gravitational corrections captured by the topological string are a 
specific specialization of the $\Omega$-deformed gauge theory, $\epo=-\ept$, it is
natural to ask whether the $\Omega$-deformed theory with general equivariant 
parameters satisfies as well a kind of anomaly equation. The main point of interest is 
that the holomorphic anomaly equation allows to analytically continue the 
$\Omega$-deformed partition function over all of the Coulomb moduli space. In contrast, the 
instanton counting partition function of \cite{N02} and as well the CFT calculations for the 
partition function via the AGT correspondence \cite{AGT09} are most useful only for the 
asymptotically free theories at a weakly coupled point in Coulomb moduli space (see however 
\cite{K12}). 

The complete partition function $Z^{\rm inst}(\epo,\ept)$ obtained via instanton
counting is exact in the two equivariant parameters $\ep_{1,2}$. In order to get started
with the investigation of the anomaly equation, one has to choose a parameterization
of the infinitesimal neighborhood of $\ep_{1,2}=0$, and the form of the answer will 
naively depend on the choice. It turns out that the correct expansion from the topological 
string point of view is to choose the same parameterization as occurring in the AGT 
correspondence, used in this context in \cite{KW10a,KW10b}. Namely, we write
\beq
\eqlabel{epdef}
\ep_1=\sqrt{\beta}g_s\,,\,\,\,\,\ep_2=-\frac{1}{\sqrt{\beta}}g_s\,,
\eq
with $\beta$ a fixed constant and $g_s$ being the only infinitesimal expansion parameter. 
Hence (leaving the Coulomb moduli $a$ implicit) we define the perturbative amplitudes 
$\Fcal^{(g)}(\beta)$ as the coefficients in the expansion
\beq
\eqlabel{Fdef}
\log Z^{\rm inst}(\ep_1,\ep_2)=
\Fcal(\ep_1,\ep_2)=\sum_{g=0}^{\infty}\Fcal^{(g)}(\beta)\, g_s^{2g-2}\,.
\eq
We note in particular that the Seiberg-Witten prepotential defined via the limit 
\eqref{limit}, is {\it independent} of $\beta$, \ie,
\begin{equation}
\Fcal(\ep_1,\ep_2)=\Fcal^{(0)} g_s^{-2}+\Ocal(g_s^0)\,.
\end{equation}
Of course, one might also envisage a double expansion in the two-parameters $\ep_1,\ep_2$,
as performed in this context in \cite{HK10}. However, it is not hard to see that
the two-parameter expansion is related via a finite resummation to the one-parameter
expansion \eqref{Fdef}. This is related to the fact that the $\Fcal^{(g)}(\beta)$ are
{\it polynomial} in $\beta$. As a consequence, an anomaly equation for a two-parameter
expansion scheme is algebraically equivalent with the anomaly for the one-parameter
expansion (\cf, the discussions in \cite{KS11,P12}). Our results, and specifically, 
the fact that only the holomorphic ambiguity depends on $\beta$, make it clear that
the one-parameter expansion \eqref{Fdef} is most economical and therefore preferred.%
\footnote{This is not to say that the microscopic origin of the holomorphic
anomaly might not be better explained in the two-parameter scheme, see \cite{P12}.}


Note that with the four-dimensional Lorentz invariance, the expansion \req{Fdef} goes 
in even powers of $g_s$ only, reflecting the symmetry $\epsilon_{1,2}\to-\epsilon_{1,2}$ 
of the $\Omega$-background. As noted in \cite{N02,okuda}, localization in the presence 
of mass parameters in principle can violate this symmetry, and odd powers of
$g_s$ will be present as well. However, this odd sector is not fundamental, and
can be ``gauged away'' by a linear shift of the appropriate mass parameters.
Notably, this does not apply to the theories in the presence of additional
extended objects like surface operators (discussed in chapter \cite{V8}). 
Such a setup breaks 4-d Lorentz
invariance and a true odd sector in $g_s$ will be generated. We will here only
consider the even in $g_s$ case, while emphasizing that the general case can
be treated as well \cite{KW10a}, by appealing to the extended holomorphic anomaly 
equations of \cite{W07a,W07b}.

So let us now explain in detail the method of the holomorphic anomaly for the
calculation of the $\Fcal^{(g)}(\beta)$. We begin with the role of $\Fcal^{(0)}$
in special geometry. We denote by $u$ a global coordinate on the moduli space 
$\calm$ of vacua, which is identified with the base space of an appropriate family 
of complex curves, $\calc_u$. (For simplicity, we will write equations only
in the case that $\calm$ is one-dimensional. The reader might have in mind
$SU(2)$ Seiberg-Witten theory with $N_f<4$ fundamental flavors. Some
aspects of the higher-rank theory are discussed in \cite{HK10,HK11}.) The family 
of curves is equipped with a meromorphic one-form $\lambda_{\rm SW}$, such that 
for appropriate choice of one-cycles $A$ and $A_D$ on $\calc_u$, the periods
\begin{equation}
\eqlabel{periods}
a = \oint_A \lambda_{\rm SW} \,,\qquad a_D = \oint_{A_D} \lambda_{\rm SW}\,, 
\end{equation}
satisfy the relation
\begin{equation}\eqlabel{SpecialGeoRelation}
a_D = \frac{\del\calf^{(0)}}{\del a} \,,
\end{equation}
after eliminating $u$ from \eqref{periods}. We do not need to be explicit
about this auxiliary geometric data, for which we refer to chapter \cite{V1}. However, 
one should keep in mind, as already mentioned in the previous section, that this 
auxiliary data originates from the mirror Calabi-Yau geometry of the corresponding 
geometric engineering geometry.

For expansion in different regions of moduli space, it is most convenient to 
base the development on the Picard-Fuchs equation, a third order 
system of linear differential equations,
\begin{equation}
\eqlabel{PFeqs}
\call \varpi(u) = 0 \,,
\end{equation}
satisfied by all periods of $\lambda_{\rm SW}$. Using $a$ as a local coordinate around 
$u\to\infty$, the Picard-Fuchs operator takes the form\footnote{As is now evident, 
the constant is a third solution of the differential equation. This solution decouples 
in special cases, such as $SU(2)$ gauge theory with massless hypermultiplets.}
\begin{equation}
\eqlabel{flatPF}
\call = \del_a\frac{1}{C_{aaa}} \del_a^2 \,,
\end{equation}
where
\begin{equation}
C_{aaa} = \del_a^3 \calf^{(0)} = \del^2_a a_D(a) = \del_a \tau(a)\,,
\end{equation}
is a (meromorphic) rank three symmetric tensor over $\calm$, which in the topological
string context is referred to as the Yukawa coupling, and $\tau(a)$ corresponds to the 
complexified effective gauge coupling. In particular, $g\sim {\rm Im}\tau$ is the 
Weil-Petersson (or $\sigma$-model) metric on $\calm$, which plays a central role in 
special geometry on $\calm$. Another important feature is the existence of canonical 
(flat) coordinates \cite{BCOV93b}, which provide a meaningful expansion parameter around any 
interesting point $u=u_*$ in $\calm$. In such a flat coordinate $t=t(u)$, vanishing 
at $u=u_*$, the Picard-Fuchs operator takes again the form \eqref{flatPF} with $a\to t$, 
\ie,
\begin{equation}
\eqlabel{general}
\call=\del_t\frac{1}{C_{ttt}}\del_t^2\,,\qquad
C_{ttt} = \Bigl(\frac{\del u}{\del t}\Bigr)^3 C_{uuu} \,.
\end{equation}

We are now ready to write down the holomorphic anomaly equations of \cite{BCOV93b}.
Recall that the specialization of the gauge theory amplitude $\calf^{(g)}(\beta)$ to 
$\beta=1$, (namely, the self-dual background $\epo=-\ept$) arises via geometric engineering 
from the genus-$g$ topological string amplitude. The statement of BCOV is that the 
topological string amplitudes, while holomorphic in the K\"ahler moduli, are not 
well-behaved globally over the moduli space. Instead, one should view the topological 
string amplitudes as a holomorphic limit of non-holomorphic, but globally defined objects. 
Under the gauge theory limit sketched in the previous section, this translates to the 
statement that one should view the gauge theory $\calf^{(g)}(a)$ (for $g\ge 1$) as the 
holomorphic limit $\bar a\to\infty$ of {\it non-holomorphic, but globally defined} 
objects $\calf^{(g)}(u,\bar u)$, arising from the topological string amplitudes in 
the gauge theory limit. (These are customarily denoted by the same letter, as 
confusion can not arise.) Similarly, the holomorphic anomaly equation satisfied by 
the topological string amplitudes translates to a recursive relation for the gauge 
theory $\calf^{(g>1)}(u,\bar u)$, \ie,
\begin{equation}
\eqlabel{hae}
\delbar_{\bar u} \calf^{(g)} = \frac 12  \sum_{\topa{g_1+g_2=g}{g_i>0}}
{\bar C}_{\bar u}^{\; u u}\calf^{(g_1)}_u \calf^{(g_2)}_u  + 
\frac12{\bar C}_{\bar u}^{\; u u} \calf^{(g-1)}_{uu} \,,
\end{equation}
where $\calf^{(g)}_{uu}=D_u \calf^{(g)}_u = D_u^2 \calf^{(g)}$, $D_u$ is the covariant 
derivative over $\calm$, and indices are raised and lowered using the Weil-Petersson 
metric. The holomorphic limit of the connection of the Weil-Petersson metric (entering 
the covariant derivative) on $\calm$ takes the simple form
\begin{equation}
\eqlabel{relation}
\lim_{\bar t\to 0} \Gamma^{u}_{uu} = \del_u \log \frac{\del t(u)}{\del u} \,.
\end{equation}
The ``one-loop'' amplitude satisfies the special equation
\begin{equation}
\eqlabel{oneloop}
\delbar_{\bar u}\del_u \calf^{(1)} =  \frac 12 {\bar C}_{\bar u}^{\; uu} C_{uuu} \,.
\end{equation}
At the level of the topological string, the holomorphic anomaly originates from topological 
anomalies, that is, under coupling to gravity (including integration over moduli of 
Riemann surfaces), the theory is only ``almost" topological, as there are contributions 
from the boundaries of moduli spaces of genus-$g$ Riemann surfaces to certain 
topologically trivial correlator insertions. More specifically, a genus-$g$ Riemann 
surface can degenerate either to a genus-$(g-1)$ surface with two extra punctures via 
pinching of a handle, or to two disconnected surfaces with an extra puncture of genus 
$g_a$ and $g_b$ (with $g=g_a+g_b$) via pinching of a tube. These two boundary contributions 
are reflected in the holomorphic anomaly equation \req{hae}. Although the topological 
string origin of the anomaly equation (and in particular of $\Fcal^{(g)}(u,\bar u)$) 
is clear, less so is the precise supergravity (or gauge theory) meaning and/or origin 
thereof.  Hence, the main justification of \req{hae} at the level of gauge theory 
comes as a limit of the topological string via geometric engineering. However,
an independent justification can be given along the lines of Witten's wave-function
interpretation of the topological string partition function \cite{W93}. The
reasoning leading to this interpretation of the $\calf^{(g)}$ and the recursive
relations between them relies solely on the special geometry (the holomorphic
symplectic structure) of the moduli space (viewed as a clasical phase space). 
Starting from the Seiberg-Witten geometry, this reasoning can therefore also be 
applied directly to the gauge theory. We will come back to this interpretation in 
section \ref{wavefunctionintsec}.

\bigskip
The natural question to ask is how \eqref{hae} should be modified away from $\beta=1$.
The answer provided by \cite{KW10a} is the simplest possible {\it not at all!} More
precisely, in \cite{KW10a}, the use of the localization formulas of \cite{N02}
resulted in the presence of non-vanishing terms of odd order in $g_s$, suggesting
a role for the extended holomorphic anomaly equation of \cite{W07a,W07b}, as well
as a relation to topological string orientifolds. The extension data (the term at order 
$g_s^{-1}$) was also identified in simple
geometric terms on the Seiberg-Witten curve. In \cite{KW10b} it was observed that the 
shift of the mass parameters \cite{okuda} removes those odd terms, as mentioned above. 
While it is reassuring to see that the formalism works well with either prescription,
we here only present the shifted version, as it is more economical.

The content of the equations \eqref{hae}, \eqref{oneloop} is that due to the anti-holomorphic 
derivative, the $\Fcal^{(g)}$ are determined up to holomorphic terms, the so-called holomorphic 
ambiguity. The standard technique to fix this ambiguity is by taking known characteristics 
of the $\Fcal^{(g)}$ at specific points in moduli space as boundary conditions into account. 
For instance, topological string amplitudes expanded near a point in moduli space where the 
target space develops a conifold singularity show a characteristic ``gap" structure 
\cite{HK06,HKQ06} (we denote the modulus of the deformation as $t_c$)
\beq
\eqlabel{FconiExp}
\Fcal^{(g>1)}= \Psi^{(2g-2)}(1)\, t_c^{-2g+2}+\Ocal(t_c^0)\,,
\eq
with leading non-vanishing coefficients $\Psi^{(g)}(1)$ of the singular terms given by 
the free energy of the $c=1$ string at the self-dual radius $R=1$ \cite{GV95} (we use here 
a normalization different from the one usually used in the CFT context). Knowledge of the 
conifold expansion \req{FconiExp} is usually sufficient to fix the holomorphic ambiguity 
to very high genus \cite{HKQ06}, or, even to fix it completely \cite{HKR08}, depending on 
the specific model. The coefficient of the singular term in \req{FconiExp}, $\Psi^{(n)}(1)$, 
can be seen as due to integrating out a single massless hypermultiplet in the effective action 
\cite{V95} and is therefore rather universal.\footnote{It that sense, the singularity
structure (but not the regular terms) in those strong coupling regions does follow from 
a field theory computation.}
In particular, expansion of the gauge theory 
free energy near a point in moduli space with a massless monopole/dyon (hypermultiplet) 
should show the same behavior, and indeed does \cite{HK06}.

In the generalization to arbitrary $\beta\neq 1$ we then must have that the boundary 
conditions are not given by integrating out a massless hypermultiplet in an anti-selfdual
background, but rather in the $\Omega$-background. Hence, the coefficients $\Psi^{(g)}(1)$ 
change to $\beta$-dependent functions $\Psi^{(n)}(\beta)$ captured by the Schwinger 
type integral
$$
\Fcal_{c=1}(\ep_1,\ep_2;t_c):=\int\frac{ds}{s} \frac{e^{-t_c s}}{4\sinh
\left(\frac{\ep_1 s}{2}\right)\sinh\left(\frac{\ep_2 s}{2}\right)} \sim 
\dots+\sum_{n>0} \Psi^{(n)}(\beta) \left(\frac{g_s}{t_c}\right)^n\,,
$$
under the usage of \req{epdef}. Interestingly, the free energy $\Fcal_{c=1}(\ep_1,
\ep_2;t_c)$ still corresponds to the $c=1$ string free energy, albeit at general 
radius $R=\beta$. The corresponding partition function is also known as Gross-Klebanov 
partition function, and we have for the expansion coefficients the following closed 
expressions \cite{GK90} 
\beq
\eqlabel{psibeta}
\begin{split}
\Psi^{(0)}(\beta)&=-\frac{1}{24}\left(\beta+\frac{1}{\beta}\right)\,,\\
\Psi^{(n)}(\beta)&=(n-1)!\sum_{k=0}^{n+2}(-1)^k\frac{B_{k}
B_{n+2-k}}{k!(n+2-k)!}(2^{1-k}-1)(2^{k-n-1}-1)\,\beta^{k-n/2-1}\,.
\end{split}
\eq

Using the coefficients \req{psibeta} as boundary conditions for general (real) $\beta$
and analytically continuing back to the weakly coupled regime, somewhat 
surprisingly reproduces for $SU(2)$ with massless $N_f<4$ flavors the instanton 
counting results of \cite{N02} (after appropriate choice of gauge of mass parameters, 
\cf, discussion above), as first reported in \cite{KW10a, KW10b}.

Similarly, using \req{psibeta} as boundary conditions for the topological string expanded 
near a conifold singularity of specific local Calabi-Yau geometries reproduces under 
analytic continuation the refined free energy defined via the 5d instanton counting. 
However, there is one important subtlety, which is usually not explicitly mentioned in 
the literature. Namely, even for a simple Calabi-Yau like local $\P^2$ or $\P^1\times\P^1$, 
the boundary conditions \req{psibeta} alone are not sufficient to completely fix the 
holomorphic ambiguity. The actual difference comes in at 1-loop. Generally, the 1-loop 
holomorphic ambiguity
possesses not only a contribution from the conifold discriminant, but also from the 
large volume divisor. For example, the 1-loop ambiguity $a^{(1)}(\beta)$ of refined 
local $\P^2$ reads \cite{ACDKV11}
\beq
a^{(1)}(\beta)=\Psi^{(0)}(\beta)\log \Delta +\kappa(\beta) \log z\,,
\eq
with $\Delta$ parameterizing the conifold locus, \ie, $\Delta:=(1-27z)$ and 
\beq
\kappa(\beta)=-\Psi^{(0)}(\beta)-\frac{2}{3}\,.
\eq
Note that in contrast to $\Psi^{(n)}(\beta)$, we do not know how to infer $\kappa(\beta)$ 
from first principles. Rather $\kappa(\beta)$ has to be manually chosen appropriately to 
reproduce the desired 1-loop free energy model by model. 

\section{Refinement vs.\ Orbifolds}

We observed in the previous section that refinement near a conifold point in moduli 
space can be interpreted as a radius deformation of the $c=1$ string. The well-known 
duality between integer radius deformations and orbifolding suggests that at least 
locally and for integer $\beta$, refinement can also be given a more geometric 
interpretation in terms of a $\Z_\beta$ orbifold. Namely, one may view the 
refinement in the B-model (for fixed integer $\beta$) near a conifold point in 
moduli space effectively as a replacement of the conifold singularity with an 
$A_\beta$ singularity. 

There is an apparent puzzle in this proposed orbifold interpretation. In the orbifold 
case, we have only an anti-selfdual background, so how do the coefficients $\Psi(\beta)$ 
arise then? Well, the answer is relatively simple. Under the $\Z_\beta$ action, 
we do not have just one, but $\beta$ massless hypermultiplets contributing to the 
leading coefficient of \req{FconiExp}. This is reflected in the fact that we can 
decompose the above Schwinger integral representation as (\cf, \cite{GV98})
$$
\Fcal_{c=1}(\ep_1,\ep_2;t_c)=\sum_{n=0}^{\beta-1} \Fcal_{c=1}(\ep_1,-\ep_1;t_n )\,,
$$
with $t_n:=t_c-n\ep_2-(\ep_1+\ep_2)/2$. Hence, the heart of the proposed orbifold 
interpretation for integer $\beta$ lies in the fact that we can trade a single 
massless hypermultiplet in the corresponding $\Omega$-background for $\beta$ 
massless hypermultiplets in an anti-selfdual background.

It is well known that the coefficients $\Psi^{(2n-2)}(1)$ correspond to the virtual Euler 
characteristic of the moduli space of complex curves of genus $n$, and one might ask for 
a similar interpretation for general $\beta$. As observed in \cite{KW12,KS13}, up to a 
shift the coefficients for general $\beta$ in fact match with the parameterized Euler 
characteristic interpolating between the virtual Euler characteristic of the moduli space 
of real and complex curves proposed in \cite{GHJ01} (see also \cite{NC13} for a more 
detailed discussion of this correspondence). The $c=1$ string orbifold interpretation 
sketched above now allows us to conjecture a geometric interpretation of this parameterized 
Euler characteristic for integer $\beta$. Namely, it should correspond to the virtual 
Euler characteristic of genus $n$ curves with a $\Z_\beta$ action. 

These local facts lead to the important question whether there exists as well a purely 
geometric interpretation of the refined partition function at large volume. That is, 
we should ask if there exists a target space $\widehat X_\beta$ such that the refined 
free energy corresponds to the count of maps
\beq
\Sigma^{(g)}\rightarrow \widehat X_\beta\,.
\eq
Naively, one would suspect that $\widehat X_\beta$ corresponds to a free $\Z_\beta$ 
orbifold of the original Calabi-Yau $X$ (for integer $\beta$). In particular, the only 
visible effect of such a free orbifold on the level of the holomorphic anomaly equations 
would be a mere change of boundary conditions, as we observed for refinement.

Indeed, one can find for specific models and values of $\beta$ concrete proof that 
the refined partition function is dual to the usual topological string on an orbifold 
of the original geometry. The simplest example has been already given in \cite{KW10a},
where it was observed that the quotient of local $\P^1\times\P^1$ by its obvious $\Z_2$ 
symmetry equals the refined partition function on the original geometry at $\beta=2$. 
A similar observation can be made for orbifolding local $\P^2$ by its cyclic $\Z_3$ 
symmetry, which corresponds to the refined partition function at $\beta=3$ \cite{DK12b}.

One should note that the a priori undetermined function $\kappa(\beta)$ leaves the 
freedom for different analytic continuations of the $\Z_\beta$ symmetry of the conifold 
to large volume. The necessity of such an ambiguity is intuitively clear, as there 
might exist at large volume differently acting symmetries, which still yield under 
analytic continuation the same leading singular behavior at the conifold point. For 
instance, there might be differently acting $\Z_2$ orbifolds at large volume, which, 
due to the high symmetry of the conifold, all possess the same coefficients 
$\Psi^{(n)}(2)$ (the massless hypermultiplet does not care which symmetry it feels). 

In general, however, it is far from clear how, if at all, the $\Z_\beta$ symmetry of 
the conifold point in moduli space translates to the large volume regime (\ie, is 
globally preserved). In the above two examples, the correspondence between the refined
topological string for particular values of $\beta$ to the usual topological string on
a different (orbifold) background could be argued to be a consequence of the large 
global symmetry group, and therefore somewhat accidental. This still leaves open
the possibility for the existence of a new classical target space $\widehat X_\beta$, 
which could be obtained for instance in the case of $\beta$ integer as a suitable 
partial compactification of an orbifold of the conifold.

\section{Wave-function interpretation}
\label{wavefunctionintsec}

The special geometry relation \req{SpecialGeoRelation} between the flat coordinate
$a$ and the magnetic dual $a_D$ is identical to the relation between canonically
conjugate variables $(p,q)$ of a classical integrable system. In particular, 
comparison with the Hamilton-Jacobi equation $H(q,\frac{\partial S}{\partial q}) 
= 0$ shows that in this interpretation, the prepotential $\Fcal^{(0)}(a)$ should
be identified with Hamilton's principal function (or classical action) effecting
the canonical transformation to the action-angle variables. 


Consider now the full perturbative partition function $Z$ expanded as a series in 
$g_s$, as in \req{Fdef}. We have
\beq
\eqlabel{wavefunctionZ}
Z= f(a,g_s,\beta)\, e^{\frac{1}{g_s^2}\int a_D\, da}\,,
\eq
with $f(a,g_s,\beta)$ some regular series in $g_s^2$. This expansion shows that 
the partition function should be interpreted as a WKB-type wavefunction in a 
semi-classical approximation to a quantization of the original hamiltonian system. 


In this context, it is important to keep in mind that quantization is intrinsically
ambiguous, \ie, in general on cannot associate a unique quantum operator $\hat H$
to a classical Hamiltonian $H$. This is most clearly apparent in the ordering
ambiguities that plague the lifting of functions of the phase-space coordinates to
quantum-mechanical operators. While the semi-classical terms are universal, the
higher order terms in the expansion \req{wavefunctionZ} are sensitive to these
ambiguities. 


The holomorphic anomaly plays in interesting role in this so-called ``wavefunction
interpretation`` of the topological partition function. In fact, as pointed out by 
Witten \cite{W93}, the holomorphic anomaly equation simply expresses the change
of the wavefunction under a change of polarization of the underlying classical 
system (\ie, the separation of the canonical coordinates into position and momentum 
variables). Above, we considered a real polarization, but complex polarizations
are natural as well. In the sense of this wave-function interpretation, the 
topological partition function $Z$ is a representation of the true ground state 
for a particular choice of polarization.

Witten's original proposal was made for the partition function of topological strings
(for which the holomorphic anomaly was first discovered), but given the results of 
\cite{HK06,KW10a} that we have reviewed above, the interpretation is very natural 
in the context of gauge theory as well (\cf, \cite{almost}). An interesting consequence of the fact 
that the holomorphic anomaly equation is insensitive to the deformation parameter
$\beta$ is that the refined partition function corresponds to a family of quantum
states with the same semi-classical expansion. On the other hand, it remains unclear
how to determine the additional conditions that would select the topological 
partition function as the unique ground state of the system. To our knowledge,
the wavefunction interpretation has not been successfully exploited for fixing 
the holomorphic ambiguity.

Via the AGT conjecture (for which there is now substantial evidence, as reviewed 
elsewhere in this volume), quantum Liouville theory provides an answer to
the quantization problem that is in principle independent from the relation to
topological strings, and has the advantage of being algorithmic. Yet another
approach to the quantization problem are the so-called topological recursions of
Eynard-Orantin \cite{eyor}. A detailed comparison between these various schemes 
remains an interesting avenue for further research.

\section{Outlook}

We see two major open problems whose solution would constitute significant progress. 
The first, more technical in nature, 
is the direct and explicit calculation of the gauge theory partition function at strong 
coupling either via instanton counting or CFT. The results reviewed here, specifically
the holomorphic limit of the $\calf^{(g)}$ as $\bar t_D=0$, $t_D\to0$, provide a benchmark 
for such a calculation. A first indication that the 2d-4d relation holds beyond weak 
coupling has been found at hand of an explicit example recently in \cite{K12}. However, 
the general picture is far from clear. The perhaps most closely related work from a 
CFT point of view is \cite{GT12}. Ultimately, a simple state like construction, as 
in \cite{G09}, for the strongly coupled expansion would be desirable.

The second major open problem is to obtain a better understanding of the 
deformation parameter $\beta$ in the topological string context. The core question 
is whether the deformation really involves a new world-sheet theory (for instance,
with a second string coupling, in case the extra parameter is viewed as an infinitesimal
coupling constant), or whether it might be sufficient to view the $\beta$-degree
of freedom entirely as a geometric deformation of the target space of the usual 
topological string. In this review, we somewhat focussed on the latter point of view.

Though we did not discuss them in this review, it has to be mentioned that there 
are several explicit proposals in the literature \cite{Antoniadisetal,NO11,P12,HK11}, 
supporting the possibility for an actual world-sheet interpretation with two infinitesimal 
coupling constants. In the mathematical formulation of the perturbative 
topological string (Gromov-Witten theory), this might involve a sort of refined count
of holomorphic maps \cite{HIV,P12}
\begin{equation}
\Sigma^{(g_1,g_2)}\rightarrow X\,,
\end{equation}
with worldsheets $\Sigma$ of genus $g=g_1+g_2$, carrying an additional $\Z_2$ 
valued decoration of the handles. 

Finally, another proposed interpretation involves replacing the B-model geometry 
by a sort of ``quantum geometry'' $\widetilde Y_q$ (encoding one of the parameters in 
a suitable parameterization).  In this approach the tree-level special 
geometry depends explicitly on the extra parameter \cite{MM09,ACDKV11}. This is 
analogous to the replacement of the spectral curve with a ``quantum'' spectral curve in the 
$\beta$-deformed matrix models reviewed in chapter \cite{V7}, and the ``quantization" 
of Seiberg-Witten theory outlined in chapter \cite{V9}. 

One should note that the quantum geometry $\widetilde Y_q$ is not directly related 
to the wave-function interpretation of the partition function discussed in section 
\ref{wavefunctionintsec} (at least the precise relation is not known). While in 
the former we quantize the underlying curve, in the latter we quantize the periods. 
In this sense we can understand the refined topological string also as a double 
quantization. For $\widetilde Y_q$ the ordinary special geometry relation is lifted 
to a so-called quantum special geometry relation and the ordinary periods to quantum 
periods (depending on the extra parameter). Quantizing similarly as in section \ref{wavefunctionintsec} the 
quantum periods, will again lead to the holomorphic anomaly equation, now for the 
(double) quantum states.

While neither proposal is entirely convincing in the present form, a possible
connection between the two proposed target space deformations might even be more 
tantalizing. Optimistically, it might indicate a new type of classical-quantum 
duality relating the topological string on $\widehat X_\beta$ with that on 
$\widetilde Y_q$. If in addition a refined topological string exists, in the 
sense of a deformed world-sheet theory, this duality would extend to a triality. 
This, at least, is the inspiration that we take away.

\begin{acknowledgments}
We would like to thank J.\ Teschner for the invitation to participate in this
joint review effort, his hard work and patience. We thank all other contributors
for their valuable comments and input.
The work of D.K. has been supported in part by a Simons fellowship, the Berkeley 
Center for Theoretical Physics and the National Research Foundation of Korea 
Grant No. 2012R1A2A2A02046739. 
The research of J.W.\ is supported in part by an NSERC discovery grant and a
Tier II Canada Research Chair. 
\end{acknowledgments}

\end{document}